\documentclass{emulateapj}
\usepackage{graphicx}
\usepackage{lipsum}

\makeatletter
\newcommand*{\rom}[1]{\expandafter\@slowromancap\romannumeral #1@}

\usepackage[backref,breaklinks,colorlinks,citecolor=black]{hyperref}
\newcommand{\OIII}{\hbox{[{\rm O}\kern 0.1em{\sc iii}]}}
\usepackage{todonotes}
\usepackage{enumitem}
\usepackage{amsmath}
\makeatother
\shorttitle{Distinguishing Mergers and Disks at $z\sim2$}
\shortauthors{Simons et al.}
\usepackage{verbatim}
\usepackage{natbib}

\begin{document}
\title{Distinguishing Mergers and Disks in High Redshift Observations of Galaxy Kinematics}
\author{Raymond C. Simons\altaffilmark{1,2}, Susan A. Kassin\altaffilmark{1,2}, Gregory F. Snyder\altaffilmark{2}, Joel R. Primack\altaffilmark{3,4}, Daniel Ceverino\altaffilmark{5, 6}, Avishai Dekel\altaffilmark{7,1}, Christopher C. Hayward\altaffilmark{8}, Nir Mandelker\altaffilmark{9,10}, Kameswara Bharadwaj Mantha\altaffilmark{11}, Camilla Pacifici\altaffilmark{2}, Alexander de la Vega\altaffilmark{1}, and Weichen Wang\altaffilmark{1}}

\affil{$^1$Johns Hopkins University, Baltimore, MD, 21218, USA; rsimons@jhu.edu\\
	$^2$Space Telescope Science Institute, 3700 San Martin Drive, Baltimore, MD, 21218, USA\\
	$^3$Physics Department, University of California, Santa Cruz, CA, 95064, USA\\
	$^4$Santa Cruz Institute for Particle Physics, University of California, Santa Cruz, CA 95064, USA\\
	$^5$Universit{\"a}t Heidelberg, Zentrum fur Astronomie, Institut f{\"u}r Theoretische Astrophysik, 69120, Heidelberg, Germany\\
	$^6$Cosmic Dawn Center (DAWN), University of Copenhagen, Lyngbyvej, 2, 2100, Copenhagen \O, Denmark\\
	$^7$Centre for Astrophysics and Planetary Science, Racah Institute of Physics, The Hebrew University, Jerusalem, 91904, Israel\\
	$^8$Center for Computational Astrophysics, Flatiron Institute, New York, NY, 10010, USA\\
	$^9$Department of Astronomy, Yale University, New Haven, CT, 06520, USA\\
	$^{10}$Heidelberg Institute for Theoretical Studies, Heidelberg,  69118, Germany\\
	$^{11}$Department of Physics and Astronomy, University of Missouri-Kansas City, Kansas City, MO 64110, USA
	}

\begin{abstract}
The majority of massive star-forming galaxies at $z\sim2$ have velocity gradients suggestive of rotation, in addition to large amounts of disordered motions. In this paper, we demonstrate that it is challenging to distinguish the regular rotation of a disk galaxy from the orbital motions of merging galaxies with seeing-limited data. However, the merger fractions at $z\sim2$ are likely too low for this to have a large effect on measurements of disk fractions. To determine how often mergers pass for disks, we look to galaxy formation simulations. We analyze $\sim$24000 synthetic images and kinematic maps of 31 high-resolution simulations of isolated galaxies and mergers at $z\sim2$. We determine if the synthetic observations pass criteria commonly used to identify disk galaxies, and whether the results are consistent with their intrinsic dynamical states. Galaxies that are intrinsically mergers pass the disk criteria for anywhere from 0 to 100$\%$ of sightlines. The exact percentage depends strongly on the specific disk criteria adopted, and weakly on the separation of the merging galaxies. Therefore, one cannot tell with certainty whether observations of an individual galaxy indicate a merger or a disk. To estimate the fraction of mergers passing as disks in current kinematics samples, we combine the probability that a merger will pass as a disk with theoretical merger fractions from a cosmological simulation. Taking the latter at face-value, the observed disk fractions are overestimated by small amounts: at most by $5\%$ at high stellar mass ($10^{10-11}$ M$_{\odot}$)  and $15\%$ at low stellar mass ($10^{9-10}$ M$_{\odot}$). 
\end{abstract}

\keywords{galaxies: evolution - galaxies: formation -galaxies: fundamental parameters - galaxies: kinematics and dynamics}

\section{Introduction}

The cosmic star-formation rate density peaks between $1\,<\,z\,<\,3$ and marks a critical period for galaxy assembly \citep{2014ARA&A..52..415M}. The processes thought to be most relevant for regulating the mass growth and structural transformation of galaxies at this time, e.g., stellar and AGN feedback, violent disk instabilities, cold mode accretion, minor/major mergers (see e.g., review by \citealt{2015ARA&A..53...51S}, and references therein), may also be dynamically disruptive, destroying ordered disk rotation on timescales comparable to the dynamical time of the galaxy.

The observed fraction of high redshift galaxies that are disk-like thus provides important insight into the frequency and relevance of these processes. If the fraction is high, it implies that early galaxy assembly is governed by dynamically calm processes, ones that are hospitable for disk formation and survival. If the fraction is low, it implies that early galaxy assembly is dominated by dynamically disruptive processes, those that tend to stall the formation and settling of well-ordered disks.

The internal kinematics of galaxies offer the most relevant observable for determining if a galaxy is a disk. A disk galaxy will {\emph{at least}} have more ordered motions than random disordered motions, as quantified by the rotation velocity $V_{rot}$ and the gas velocity dispersion $\sigma_g$, respectively. 

Emission-line kinematics of the ionized gas in galaxies have now been measured for several hundred star-forming galaxies at $1 \lesssim z \lesssim 3$ (e.g., \citealt{2003ApJ...591..101E, 2006Natur.442..786G, 2006ApJ...645.1062F, 2006ApJ...646..107E, 2007ApJ...669..929L, 2007ApJ...660L..35K, 2007ApJ...658...78W, 2008ApJ...687...59G, 2008A&A...477..789Y, 2008A&A...484..173P, 2008ApJ...682..231S, 2009ApJ...697.2057L, 2009ApJ...706.1364F, 2010MNRAS.402.2291L, 2010MNRAS.404.1247J,  2011A&A...528A..88G, 2012MNRAS.426..935S,2012Natur.487..338L, 2012ApJ...758..106K, 2012A&A...539A..91C, 2013ApJ...767..104N, 2013PASA...30...56G, 2015ApJ...799..209W, 2016ApJ...831...78M, 2016ApJ...819...80P, 2016ApJ...827...57O, 2016ApJ...830...14S, 2016A&A...591A..49C, 2016MNRAS.457.1888S, 2017MNRAS.465.1157R, 2017ApJ...838...14M, 2017ApJ...839...57S, 2017MNRAS.471.1280T, 2017ApJ...842..121U, 2018arXiv180908236S}) --- with samples that are large and representative enough to derive meaningful population statistics.

At $z\,=\,2$, approximately 50\% of low mass galaxies (M$_*=10^{9-10}$ M$_{\odot}$) and 70\% of high mass galaxies (M$_*=10^{10-11}$ M$_{\odot}$) have $V_{rot}/\sigma_g\,>\,1$, i.e., more rotational support than dispersion support (\citealt{2017ApJ...843...46S}). Integral field spectroscopy (IFS) allows one to check for additional disk signatures in two-dimensional velocity and velocity dispersion maps. For the KMOS$^{3D}$ IFS survey, \citet{2015ApJ...799..209W} define a series of increasingly strict criteria that are used to identify disk-like systems. These disk criteria include: a continuous single gradient in the velocity map, a rotation velocity that exceeds the velocity dispersion, a maximum in the velocity dispersion map that is coincident with the dynamical center, an alignment between the photometric and kinematic major axes, and a spatial coincidence of the continuum and dynamical centers.  Using these, they conclude that $\sim$50 - 70\% of massive galaxies at $z\sim2$ are disk-like, with the exact fraction depending on the number of criteria used \citep{2015ApJ...799..209W}.

However, these observed kinematic signatures of disks may not be {\emph{unique}} to disks. The main challenge comes from mergers. The orbital motions of merging galaxies, once convolved with the typical observational seeing-limit at $z=2$ ($\sim0.\arcsec6$ or 5 kpc), can mimic the regular ordered rotation of a disk.  While techniques have been developed for distinguishing mergers and disks in kinematics observations (e.g., {\tt{Kinemetry}}, \citealt{2006MNRAS.366..787K}),  they are typically restricted to a small subset of observations at $z\sim2$ with sufficient resolution and signal-to-noise (e.g., \citealt{2008ApJ...682..231S}).  Using a set of artificially-redshifted local mergers and synthetic observations of idealized binary merger simulations, \citet{2015ApJ...803...62H, 2016ApJ...816...99H} demonstrate that the ability to distinguish between disks and mergers in kinematic data depends strongly on the interaction phase, becoming progressively more difficult during later stages.

Merger fractions are found to increase from $z=0$ to $z=2$ in both observations and simulations \citep{2011ApJ...742..103L, 2015MNRAS.449...49R, 2018MNRAS.475.1549M}. This means that the confusion of mergers as disks is likely more important to the observed disk fractions at higher redshifts. \citet{2017MNRAS.465.1157R} re-analyzed the \citet{2015ApJ...799..209W} sample at $z\sim1$ and conclude that 58$\%$ are involved in a merger (from first approach to post-collision) and that only one-third are isolated and virialized disks. It is clear that mergers present a potentially important problem for determining disk fractions in seeing-limited kinematics observations at high redshift, but how significant this problem is remains an open question.

In this paper, we quantify the significance of merger contamination in current high redshift seeing-limited kinematic surveys using state of the art galaxy formation simulations. First, we create and analyze synthetic observations of a suite of high-resolution zoom-in hydrodynamic simulations \citep{2014MNRAS.442.1545C} to determine the probability that a merger in a given stage will pass as a disk in the observations. These simulations are run in a cosmological context and include realistic cosmic accretion and galaxy-galaxy mergers. We then use theoretically-derived merger fractions from the large Illustris cosmological simulation \citep{2014MNRAS.445..175G, 2014MNRAS.444.1518V} to determine if there are enough galaxies in mergers (at all stages) to significantly affect global disk fractions at $z\sim2$.

The paper is structured as follows. In \S2 we describe the simulation suite. In \S3 we characterize the intrinsic dynamical properties of the simulated galaxies and identify mergers and close pairs. In \S4 we use the {\tt{SUNRISE}} dust-radiative transfer program to generate synthetic observations of the simulations. In \S5, we measure photometric properties from the synthetic imaging and kinematic properties from the synthetic spectra. In \S6 we compare the interpretation from the synthetic data with the intrinsic dynamical state of each galaxy and determine the probability that a merger will pass as a disk as a function of projected separation. In \S7 we use merger fractions from the Illustris simulation to determine if merger contamination plays a significant role in current global disk fraction statistics at $z\,=\,2$. In \S8 we summarize our conclusions. We adopt a $\Lambda$CDM cosmology defined with ($h$, $\Omega_m$, $\Omega_{\Lambda}$) = (0.7, 0.27, 0.73).

\section{VELA Simulations}

We analyze a set of 31 cosmological zoom-in galaxy formation simulations from the VELA simulation suite \citep{2014MNRAS.442.1545C, 2015MNRAS.450.2327Z}. The simulations are briefly reviewed here and we refer to \citet{2014MNRAS.442.1545C} for further details.

The simulations were run with the {\emph{N}}-body and Eulerian gas dynamics Adaptive Refinement Tree code ({\tt{ART}}, \citealt{1997ApJS..111...73K}), including subgrid recipes for: gas and metal cooling, UV-background photoionization, stochastic star formation, gas recycling and metal enrichment from stellar winds, thermal feedback from supernovae and feedback from young stars through radiation pressure and radiative heating \citep{2010MNRAS.404.2151C, 2012MNRAS.420.3490C, 2014MNRAS.442.1545C}. The adaptive mesh grid has a maximum resolution between 17 and 35 physical parsecs and this is typically reached for gas densities between $\sim10^{-2}$ -- $100$ cm$^{-3}$. The mass of the star particles ranges from $10^3$ to $10^6$ M$_{\odot}$, depending on the gas mass of the parent cell, with a typical value of $10^4$ M$_{\odot}$.

The simulated galaxies have stellar masses spanning $9.3<\log$ M$_{*}$/M$_{\odot}<10.7$ at $z = 2$ --- comparable to the typical stellar mass range of seeing-limited kinematics samples in the literature at this redshift (e.g., \citealt{2015ApJ...799..209W, 2016ApJ...830...14S}). The VELA simulations have been used in the past to study the assembly and transformation of high redshift galaxies, in particular the origin and evolution of stellar elongation \citep{2015MNRAS.453..408C, 2016MNRAS.458.4477T}, clumps \citep{2014MNRAS.444.1389M, 2016MNRAS.456.2052I, 2017MNRAS.464..635M}, morphologies \citep{2015MNRAS.451.4290S}, inflows and outflows \citep{2016MNRAS.460.2731C, 2016MNRAS.457.2605C}, galaxy-halo spin correlations \citep{2018arXiv180407306J}, and compact galaxy structure \citep{2015MNRAS.450.2327Z, 2016MNRAS.458..242T, 2016MNRAS.457.2790T, 2018ApJ...858..114H}.

The subset of independent simulations analyzed in this paper span $1\,<\,z\,<\,3$ and are separated in equal intervals of cosmological scale factor $\Delta a$ = 0.01. This corresponds to time intervals of 110 Myr at $z=3$, 125 Myr at $z = 2.0$ and 160 Myr at $z=1.0$ and totals over 1000 simulation snapshots.

\begin{figure*}
\begin{centering}
\includegraphics[angle=0,scale=0.85]{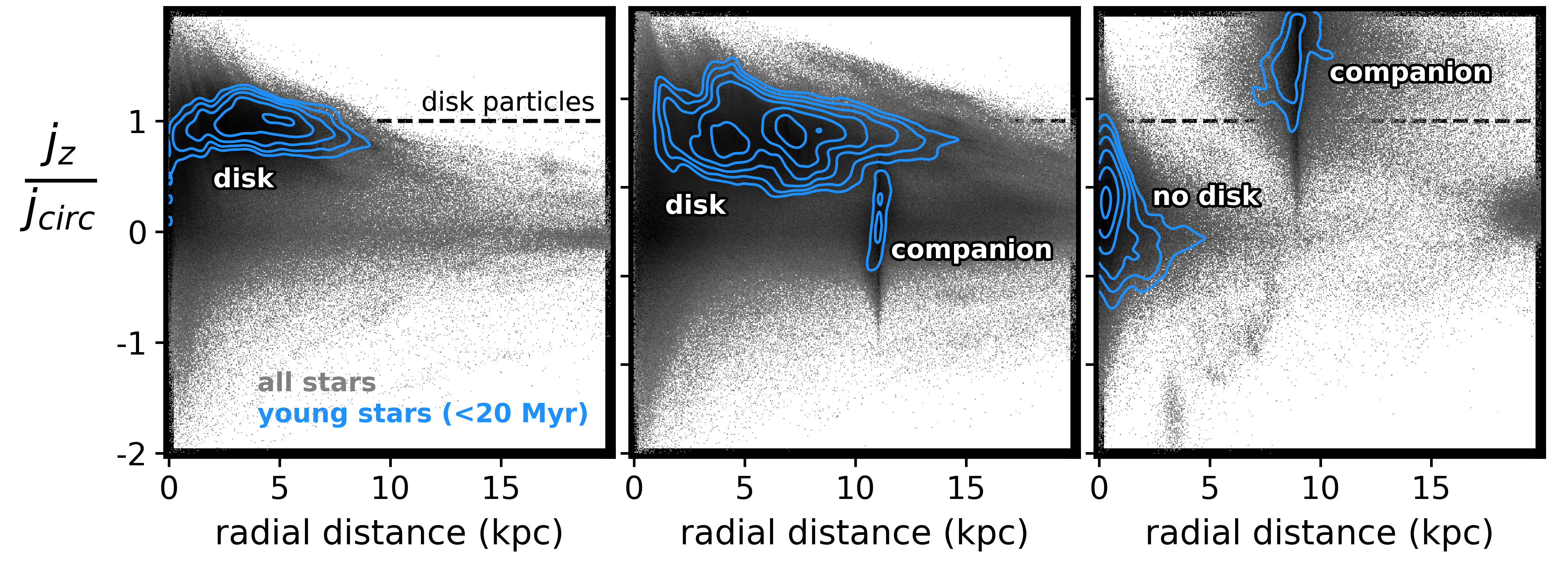}
\caption{The intrinsic states of the simulated galaxies are determined at each snapshot from their young star particles (ages $<20$ Myr). The distribution of star particles in $j_z$/j$_{circ}$ and radial distance from the galaxy centers is shown. These distributions allow one to identify disks, companion galaxies, and non-disk galaxies. The 3 panels show: a galaxy with a young stellar disk (left), a galaxy with a young stellar disk and a companion (middle), and a galaxy without a young stellar disk but with a companion (right). Particles in a disk lie near $j_z/j_{circ}=1$ (dashed black line), i.e., on circular orbits that are co-rotational with the net angular momentum of the galaxy. Grey points show all the star particles and the blue contours show the young star particles.}
\label{fig:images_intrinsic1}
\end{centering}
\end{figure*}

\section{Intrinsic States of the Simulated Galaxies}

The objective of this paper is to compare the {\emph{intrinsic properties}} of the simulations with the interpretation one would reach from their synthetic observations. The intrinsic properties of the galaxies are measured directly from the simulation data. For each simulation snapshot (i.e, time step), we use the three-dimensional positions and velocities of the star particles to (i.) determine if the most massive galaxy in the simulation box, hereafter referred to as the {\it central galaxy}, has a rotationally-supported star-forming disk and (ii.) identify all galaxies that are nearby the central galaxy, hereafter referred to as {\it companion galaxies}. As follows, the central galaxy in each simulation snapshot is categorized into one of the following: isolated disk galaxy, disk galaxy with a companion, isolated non-disk galaxy, or a non-disk galaxy with a companion.

\subsection{Using Intrinsic Properties of the Simulated Galaxies to Identify Disks}

In this section, we determine whether the central galaxy of each simulation snapshot has an intrinsic disk. Galaxy kinematics at high redshift are almost exclusively measured from emission lines, tracing the ionized gas in the star-forming regions of galaxies. For an appropriate comparison to the observations, we use the young star particles ($<20$ Myr) to search for disks in the simulated galaxies.

In a disk galaxy, stars will tend to have orbital motions aligned with the net galaxy rotation, while in a dispersion-supported galaxy they will have no preferred orbital direction. 

To assess the alignment of each star particle's orbit with the net galaxy rotation, we adopt the widely-used circularity parameter $j_z/j_{circ}$ (e.g., \citealt{2003ApJ...597...21A, 2005MNRAS.363.1299O, 2007MNRAS.374.1479G, 2008MNRAS.389.1137S,  2009MNRAS.396..696S, 2012MNRAS.423.1544S, 2015MNRAS.447.3291C}). The quantity $j_z$ is the component of a particle's specific angular momentum aligned with the net angular momentum of the galaxy.  The quantity $j_{circ}$ is the specific angular momentum of a theoretical circular orbit at the particle's position, irrespective of the orbital direction, and is calculated from the total mass inside of a shell at that position, $j_{circ}(r)\,=\,\, r\,v_{circ}(r)\,=\,\sqrt{r\,M(<r)G}$, where spherical symmetry is assumed for simplicity.

The direction of the net angular momentum of the central galaxy ($\hat{z}=\vec{J}_{gal}/J_{gal}$) is defined using an iterative fit of a cylindrical disk to both the cold gas ($T<1.5\times10^4$ K) and the stars with ages less than 100 Myr (see \S3.2 of \citealt{2017MNRAS.464..635M} for details). 

The circularity parameter is calculated for every star particle in the simulation box. Those particles associated with a disk lie near $j_z/j_{circ}\,=\,1$, i.e., on circular orbits aligned with the net rotation of the galaxy. In a dispersion-supported system, star particles are distributed symmetrically around $j_z/j_{circ}\,=\,0$. Particles with $j_z/j_{circ}\,=\,-1$ are traveling retrograde to the net galaxy rotation.

The joint distribution of stars in $j_z/j_{circ}$ versus distance from the central galaxy is shown for three example snapshots in Figure \ref{fig:images_intrinsic1}. External galaxies appear as distinct clusters of points in this plane, i.e., their stars are close to one another in space and are traveling on similar trajectories.

For each simulation snapshot, we create a segmentation map of the joint distribution of $j_z/j_{circ}$ versus distance. This is used to detect significant clusters of star particles. The segmentation procedure iterates until the number of detected objects does not change with an increasing clipping level. To prevent spurious detections, a threshold is set such that a companion galaxy must have a minimum total stellar mass of 10$^6$ M$_{\odot}$. Once this procedure is complete, each star particle in the simulation is associated with a unique galaxy. We record the mass and mass-weighted position of all galaxies detected within 100 kpc of the central galaxy.

For each timestep of each simulation, we calculate the mean circularity of the young star particles that are associated with the central galaxy. 

\begin{figure}
\begin{centering}
\includegraphics[angle=0,scale=0.68]{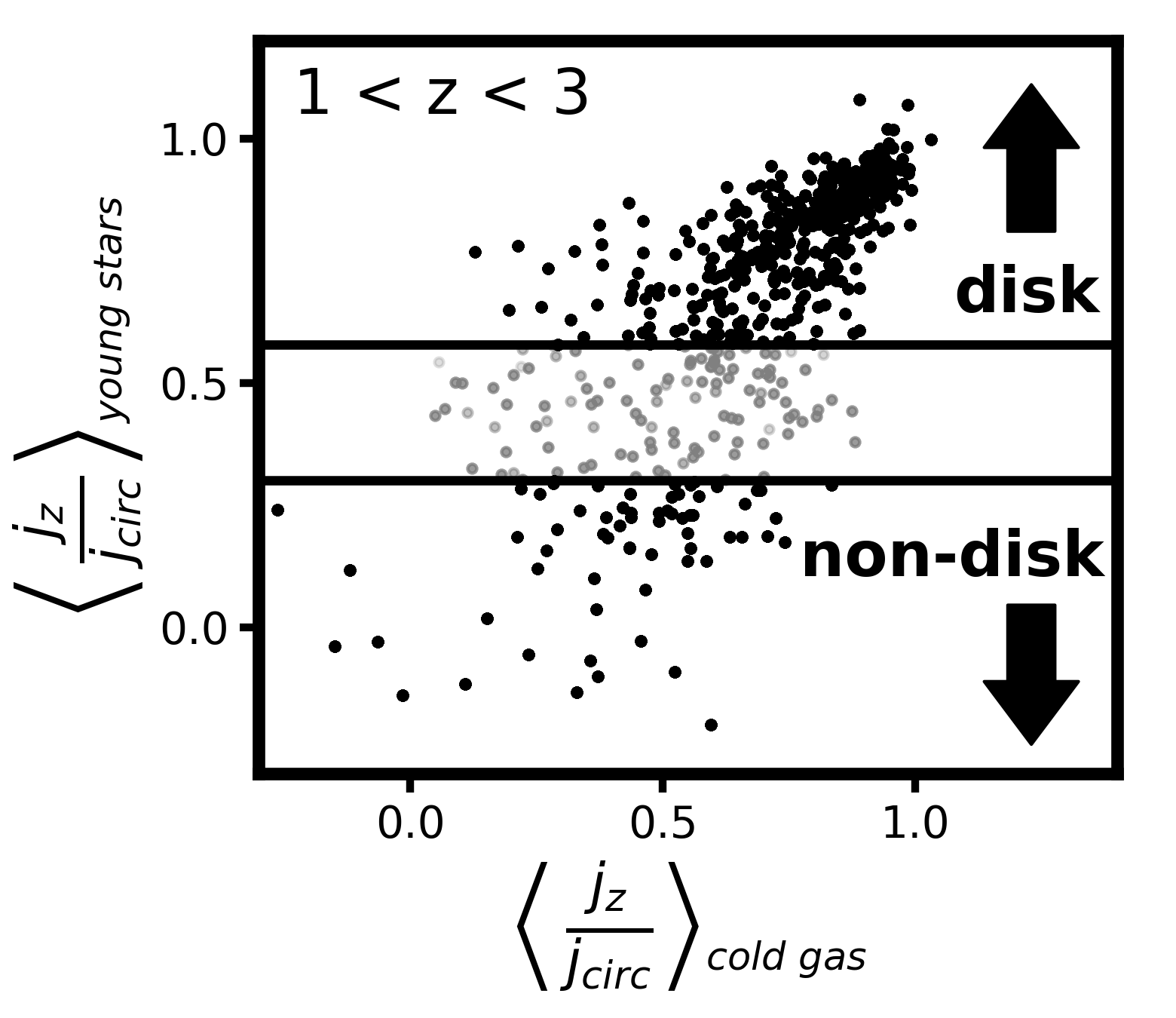}
\caption{To select simulated galaxies with and without intrinsic disks, we use the average circularity $\left<j_z/j_{circ}\right>$ of their young star particles. The average circularities of the young stars and cold gas for all galaxies considered in this paper are shown. Galaxies are defined to have an intrinsic disk if the average circularity of their young stars is greater than $1/\sqrt{3}$ (upper black line), and to not host a disk if it is less than 1/3 (lower black line). Snapshots with intermediate values are ignored (grey points). Over $1\,<\,z\,<\,3$, the majority of central galaxies in the VELA simulation have young stellar disks. The average circularities of the young stars and  cold gas correlate for the disks, but do not for the non-disks.\vspace{0.1cm}}
\label{fig:jz_jcirc_pop}
\end{centering}
\end{figure}

Each central galaxy of each simulation snapshot is determined to host a disk if the average circularity of its young stars $\left<j_z/j_{circ}\right>_{young}$ is greater than $1/\sqrt{3}$, i.e., the young stars are strongly rotationally-supported, and to not host a disk if $\left<j_z/j_{circ}\right>_{young}$ is less than 1/3, i.e., the young stars are strongly dispersion-supported (Figure \ref{fig:jz_jcirc_pop}). This classification applies only to their young stellar component --- the older stars are typically more dispersion-supported (Figure \ref{fig:images_intrinsic1}).

These thresholds in $\left<j_z/j_{circ}\right>_{young}$ are motivated by the following argument. The Jeans equation relates the circular velocity, the rotation velocity and the 1D velocity dispersion as $V_{circ}^2 = V_{rot}^2 + \alpha\,\sigma_r^2$, where $\alpha=2-3$ depending on the potential. For an isothermal sphere, $\alpha =2$ and $ j_{z}/j_{circ} >1/\sqrt3$ and $<1/3$ are equivalent to $V_{rot}/\sigma_r\,>1$ and $<$ 0.5, respectively. The former is the standard observational threshold for classifying a galaxy as a disk ---  and is one of the observational thresholds used later in this paper ---  and the latter is sufficiently low for the galaxy to be considered dispersion-dominated. For simplicity, we ignore galaxies with intermediate $\left<j_z/j_{circ}\right>_{young}$ to these two cases.

Over $1\,<\,z\,<\,3$, the majority of the central galaxies in VELA have $\left<j_z/j_{circ}\right>_{young}\,>\,1/\sqrt{3}$ --- i.e., they have a young stellar disk according to our classification (Figure \ref{fig:jz_jcirc_pop}). We stress that the goals of this paper do not rely on the true disk fraction of the VELA simulations. Our focus is to test how well one can recover the intrinsic properties of a galaxy from its synthetic observations --- an exercise that is impartial to the rarity of those intrinsic properties in the simulation set.

\subsection{Using Intrinsic Properties of the Simulated Galaxies to Identify Companion Galaxies}

As described above, we use the joint distribution of stars in radial position and $j_z/j_{circ}$ to identify sufficiently massive ($>10^{6}$ M$_{\odot}$) external galaxies in the simulation box.

Companion galaxies are defined to be those nearby the central galaxy (3D separation $\,<35$ kpc) with sufficiently high relative stellar mass (stellar mass ratio with the central less than 30). As such, we include major (1:1 - 1:4 mass ratios), minor (1:4 - 1:10), and very minor (1:10-1:30) companions. The latter two are referred to as simply `minor' for the rest of this paper. If more than one companion galaxy is detected in a snapshot, we only consider the most massive one for this paper.

A central galaxy is considered isolated if it meets the following two criteria: there are no galaxies identified within 75 kpc in the current snapshot and it did not have a companion galaxy in the previous snapshot. The second criteria ensures that we exclude galaxies that experienced a merger within the past 150 Myrs.

\section{Synthetic Observations of the Simulations: images and spectra}

To generate synthetic observations, each simulation snapshot is post-processed with the dust radiative transfer software {\tt SUNRISE} \citep{2006MNRAS.372....2J, 2010MNRAS.403...17J, 2010NewA...15..509J}. Synthetic images and spectra are created from the {\tt SUNRISE} output.

The synthetic {\emph{HST}} images used in this paper are available as a high-level science product from MAST\footnote{https://archive.stsci.edu/prepds/vela}: \dataset[https://doi.org/10.17909/t9-ge0b-jm58]{https://doi.org/10.17909/t9-ge0b-jm58}.

\subsection{Dust Radiative Transfer with {\tt SUNRISE}}

{\tt SUNRISE} contains two primary steps. It first generates a spectral energy distribution (SED) for each source of radiation in the simulation. It then propagates the associated polychromatic rays through the 3D gas grid of the simulation, taking into account dust absorption and scattering, towards a user-defined camera(s).

For each snapshot, the {\tt ART} output is first loaded into the {\tt YT} analyses software \citep{2011ApJS..192....9T} and converted into the appropriate {\tt SUNRISE} format. This is carried out using an updated version of the pipeline used in \citet{2014MNRAS.444.1389M}.

An SED is assigned to each star particle based on its age, mass, and metallicity using {\tt STARBURST99} \citep{1999ApJS..123....3L} assuming a \citet{2001MNRAS.322..231K} initial mass function. Emission lines are generated from young star particles, less than 10 Myrs old, using a starburst model from the photoionization and shock modeling code {\tt MAPPINGS III} \citep{2005ApJ...619..755D, 2008ApJS..176..438G}. This model includes a central massive young stellar cluster, whose SED is set with {\tt STARBURST99}, a surrounding HII region, and a photodissociation region with a covering fraction of 0.2 \citep{2008ApJS..176..438G}. To account for $10^4$ K thermal broadening, we convolve the original {\tt MAPPINGS III} H$\alpha$ emission line with a 10 km s$^{-1}$ 1D Gaussian kernel.

The dust density is assumed proportional to the gas metal mass using a dust-to-metals mass ratio of 0.4 (Dwek 98). Consequently, the dust geometry is resolved to the typical gas cell size of 17-35 pc. We adopt the dust grain size distribution from \citet{2001ApJ...548..296W} and \citet{2007ApJ...657..810D} and set the slope of the dust extinction law, $R_v$, to the Milky Way value of $3.1$ (e.g., \citealt{2003ApJ...594..279G}).

{\tt SUNRISE} cameras are placed at 19 positions around the galaxy. The camera parameters are user-selected and not default to SUNRISE. Of the cameras, 6 are random but remain fixed in comoving space over time, 8 are random but change from timestep to timestep, and the remaining 5 are fixed with respect to the net angular momentum of the galaxy and, as such, change from timestep to timestep. Those fixed to the galaxy include one which is face-on, i.e., along the gas angular momentum axis, one which is reverse face-on, two which are edge-on, and one which is placed at a 45$^{\circ}$ angle from face-on. Multiple sightlines allow us to overcome the intrinsic randomness associated with a single viewing angle. To avoid highly-uncertain inclination-corrections, we do not use the 2 face-on cameras or any of the random cameras that are oriented within $20^{\circ}$ of the face-on direction for a given snapshot. 

We run the radiative transfer step of {\tt SUNRISE} twice, once to create broadband images and once to create high-resolution spectra. A sufficient number of polychromatic rays are used for both runs, 10$^7$, to overcome Monte Carlo noise. 

In the first mode, {\tt SUNRISE} is run at low spectral resolution but with coverage over the NUV--NIR (0.02 -- 5 $\micron$). Each camera contains 800 x 800 pixels and covers a physical FOV of either 50 or 100 kpc. The larger FOV is used for 6 of the largest galaxies in the simulation suite to ensure sufficient coverage in their outskirts at low redshift. The output spectral cube is integrated over the spectral response function of the far-UV through IR filters available with {\emph{HST}}/ACS--WFC3, taking into account the redshift of the simulation and the corresponding cosmological Doppler shift and surface brightness dimming of the SED. 

In the second mode, {\tt SUNRISE} is run at high spectral resolution around the H$\alpha$ emission line (0.65 -- 0.66 $\micron$). The camera contains 400 x 400 pixels and covers the same physical FOV (50 or 100 kpc) as in the first mode. The pixel-by-pixel surface brightness in the resulting {\tt SUNRISE} cube is scaled with redshift to account for cosmological surface brightness dimming.

This process is repeated for each snapshot of each galaxy simulation. The final suite contains thousands of idealized UV--IR band images and idealized spectral cubes around the H$\alpha$ emission line.

\subsection{Synthetic HST/WFC3 imaging}

Following \citet{2015MNRAS.451.4290S}, the idealized images are degraded in noise and spatial resolution to simulate the typical imaging quality of modern deep-field galaxy surveys with {\emph{Hubble/WFC3}} \citep{2011ApJS..197...35G,  2011ApJS..197...36K}. The angular sizes and fluxes of the pixels are scaled with the angular size distance and luminosity distance of the simulation redshift. The images are convolved with the typical point-spread function of the instrument/filter \citep{2011SPIE.8127E..0JK} and the pixel size is rebinned to the mosaic pixel scale of the CANDELS survey (0.\arcsec06 pixel$^{-1}$ for {\emph{HST}}/WFC3-IR; \citealt{2011ApJS..197...36K}). Finally, shot noise is added to reach a specific surface brightness limit. We generate images for two surface brightness limits --- 25 mag arcsec$^{-2}$ and 27 mag arcsec$^{-2}$.

In this paper, we use the synthetic {\emph{HST}}/WFC3 F160W image (hereafter referred to as H-band) at the deeper surface brightness limit of 27 mag arcsec$^{-2}$. The observed H-band traces the rest $\sim$V-band at z = 1 and the $\sim$I-band at z = 3.

\subsection{Synthetic VLT/KMOS Integral Field Spectroscopy }

The idealized high spectral resolution cube from {\tt SUNRISE} is degraded to the surface brightness depth, spectral and spatial resolution of typical ground-based seeing-limited data. The majority of the large IFS kinematics surveys at $1\,\lesssim\,z\,\lesssim\,3$ are performed with the KMOS near-infrared spectrograph on the VLT, e.g., KMOS-3D \citep{2015ApJ...799..209W}, KROSS \citep{2016MNRAS.457.1888S}, KDS \citep{2017MNRAS.471.1280T}. For an apt comparison with the literature, we use the specifications of this instrument to generate the synthetic spectra. We emphasize, however, that our conclusions are general to all seeing-limited instruments/surveys.

The physical pixel size of the {\tt SUNRISE} cube is 0.25 kpc px$^{-1}$ and the corresponding angular scale is determined from the simulation redshift. Each spectral slice of the cube is blurred with a 2D Gaussian of 0.\arcsec6 FWHM to simulate the typical atmospheric seeing in the NIR. The pixels are rebinned to  0.\arcsec2, the angular pixel scale of KMOS, and the spectral dimension is convolved with a 1D Gaussian of $R = \Delta\lambda/\lambda = 3800$ to simulate the KMOS spectral resolution. Finally, shot noise is added to each pixel to reach a typical 5$\sigma$ point source depth of 24 AB magnitudes, which is equivalent to a typical 8 hour exposure with KMOS.

Although we aim to simulate realistic ground-based observations, we do not include sky lines and ignore that the sky transmission is significantly lower over the observed wavelengths of H$\alpha$ at $1.8\,<z\,<2.2$. These assumptions do not affect the conclusions of this paper.

\begin{figure*}
\begin{centering}
\includegraphics[angle=0,scale=1.53]{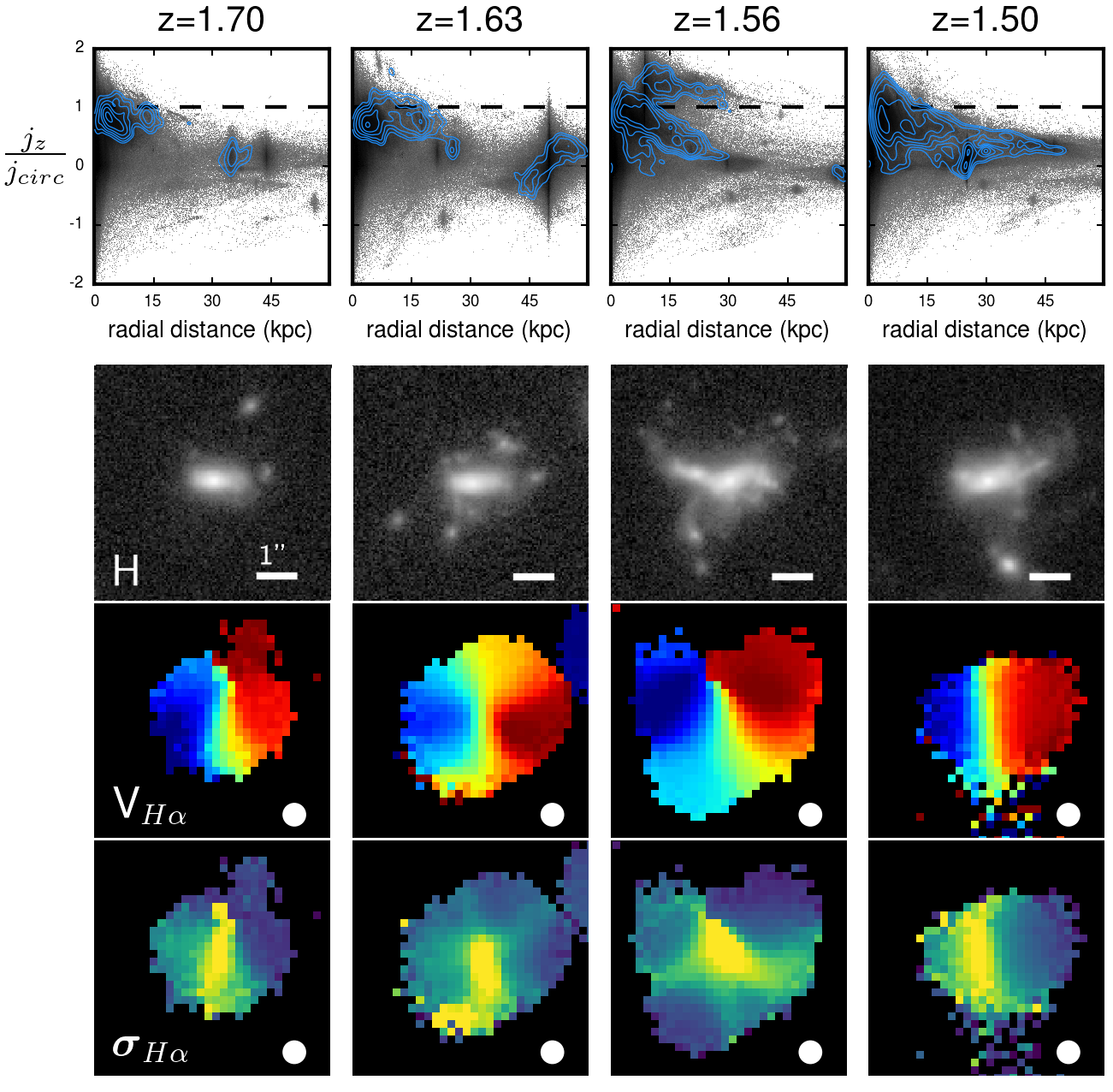}
\caption{The intrinsic states (top row) and synthetic observations (bottom rows) for 4 timesteps (columns from left to right) of a simulated major merger at $z\sim1.6$ are shown.  This galaxy is larger and better-resolved than the typical simulated mergers in our sample (as shown in later figures), and so the kinematic signatures of the merger are apparent. The central galaxy retains a disk during the merger, as indicated by the blue contours at $j_z/j_{circ}\approx1$. In the first two columns, the galaxies are sufficiently separated that they do not significantly disturb each other. The synthetic observations therefore show the regular kinematic and photometric features of a disk. In the third column, the merging galaxies are close enough to significantly disturb each other. The synthetic data therefore show more complicated features: there are two distinct gradients in the velocity map and a warp in the {\emph{Hubble}} image. In the top row, $j_z/j_{circ}$ versus radial distance is shown for all stars (black points) and young stars ($<20$ Myrs; blue contours), as in Fig. \ref{fig:images_intrinsic1}. Synthetic observations are shown in the bottom rows (from top to bottom): the {\emph{HST}}/WFC3 F160W H-band image, the VLT/KMOS velocity map, and the VLT/KMOS velocity dispersion map. The white dashed line indicates $1"$ on the sky, and the white circle indicates the FWHM of the point spread function used to create the synthetic KMOS spectral cube. The position of the camera is fixed in space from snapshot to snapshot.}
\label{fig:single_case}
\end{centering}
\end{figure*}

\begin{figure}
\begin{centering}
\includegraphics[angle=0,scale=0.34]{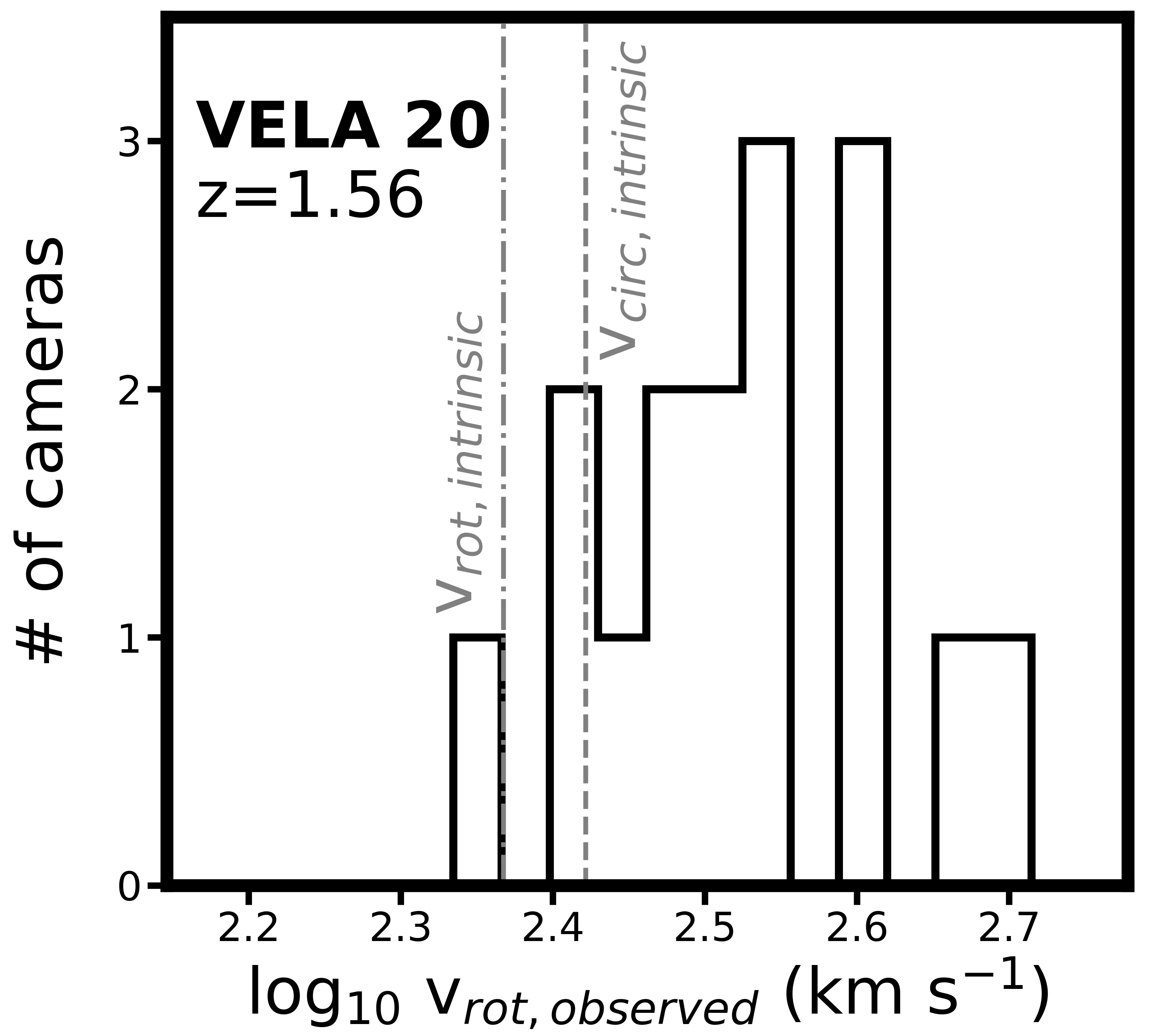}
\caption{The distribution of the observed inclination-corrected rotation velocities for 16 camera sightlines around the simulated major merger in Fig. \ref{fig:single_case} at $z=1.6$ is shown. The galaxy retains its disk during the merger. The observed rotation velocity is measured along the major kinematic axis. For most sightlines, this traces the orbital motions of the merging galaxies instead of the rotational motions of the disk. This quantity is higher than the intrinsic rotation velocity of the central disk (dashed-dotted line), and the maximum circular velocity of the galaxy-dark halo system (dashed line). The latter quantity measures the system's total dynamical mass. This demonstrates that observations of a late-stage major merger can lead to an overestimated and unphysical measurement of rotation velocity.\vspace{0.1cm}}
\label{fig:vela20_histogram}
\end{centering}
\end{figure}

\section{Measurements from the synthetic observations}

\subsection{Inclination, Photometric Position Angles, and Continuum Centroid}

Photometric properties --- inclinations and position angles --- are measured from the synthetic {\emph{HST}} H-band images. Unique sources in each image are identified and characterized using the {\tt{Photutils}} package of Astropy \citep{2013A&A...558A..33A}, with the requirement that at least 5 connected pixels have a flux that exceeds the background rms. 

Each image is centered on the central galaxy and so we select the central source as the main target. The ratio of the isophotal semi-major and semi-minor axis, $a$ and $b$, respectively, is used to determine the inclination, as $\cos^2i = ((b/a)^2 - q_0^2)/(1-q_0^2)$. We adopt a value of $q_0=0.25$, the standard literature assumption for a thick disk. The photometric position angle is defined as the direction of the semi-major axis. We determine the continuum centroid by calculating the light-weighted center of the central source in the synthetic H-band image.

\subsection{Kinematics}

The line-of-sight velocity and velocity dispersion are measured for each spaxel of the synthetic KMOS cube. To do this, we fit the H$\alpha$ emission line in the spaxel with a 1D Gaussian profile. The width, amplitude and center are left as free parameters in the fit. The center of the best-fit profile is taken as the mean line-of-sight velocity of that spaxel. The velocity dispersion of the spaxel is taken from the RMS width of the best-fit after subtracting the spectral resolution of the instrument in quadrature ($\sigma(R) = \sqrt{\sigma^2_{measured} - \sigma^2_{instrument}}$). This is repeated for all spaxels. The result of this procedure is a 2D map of the velocity and velocity dispersion. This is repeated for all of the synthetic KMOS cubes. Example kinematic maps are shown in Figure \ref{fig:single_case}.

The kinematic major axis is defined as the line intersecting the the centroid of the pixels with the maximal and minimal 10\% velocities, $V_{max}$ and $V_{min}$, respectively. A 3-spaxel wide slit is placed along the kinematic major axis. The rotation curve and velocity dispersion profile is measured along this slit by averaging the velocity and velocity dispersion of the 3 spaxels perpendicular to the slit axis.

The rotation velocity uncorrected for inclination, $V_{rot}\times\sin i$, is taken as the difference between the maximum and minimum velocities of the rotation curve. This quantity is then inclination-corrected using the inclination derived from the synthetic {\emph{HST}}/WFC3 H-band image:

\begin{equation}
V_{rot} = \frac{V_{max} - V_{min}}{2\times\sin i}
\end{equation}

The seeing will tend to smear unresolved velocity gradients and artificially elevate the velocity dispersion in the centers of galaxies, where the rotation curve is steepest (see Fig. 9 of \citealt{2006ApJ...653.1027W} and the Appendix of \citealt{2015MNRAS.452..986S}). The severity of beam smearing depends on several factors, including the intrinsic shape of the rotation curve and the ratio of the seeing to the galaxy size \citep{2016ApJ...826..214B}. To minimize the effects of beam smearing, we adopt the technique used in \citet{2015ApJ...799..209W}. We assume that the velocity dispersion is constant across the face of the galaxy and measure its intrinsic value using the average uncertainty-weighted mean of the velocity dispersion profile on the galaxy outskirts, i.e., those at and beyond the maximal and minimal 10\% velocities (typically near $\sim$1.5 $\times$ effective radius).

The dynamical center is defined along the kinematic major axis as the location where $v_{sys}=0$, i.e., the midpoint of the maximal and minimal velocities. In practice, the position is measured by interpolating between the locations of the 4 pixels that are contiguous to the pixel that is nearest to $v_{sys} = 0$ (2 on either side). The location of the peak velocity dispersion is measured from the 2D velocity dispersion map. The uncertainties on both of these quantities are calculated using a Monte Carlo technique: the values of the velocity and velocity dispersion maps are resampled by their uncertainties 1000 times and, for each iteration, the dynamical center and the location of the velocity dispersion peak are re-measured. The width of the resulting distributions of measurements is taken as the uncertainty.

\subsection{Criteria for classifying a galaxy as a disk}

To determine whether the synthetic observations are consistent with a disk-like system, we adopt the disk criteria outlined in \citet{2015ApJ...799..209W}. The first three criteria are:

\begin{enumerate}
\item The velocity map exhibits a continuous velocity gradient along a single axis.
\item The inclination-corrected rotation velocity exceeds the ionized gas velocity dispersion, i.e., $V_{rot}$/$\sigma_{{\mathrm{H}\alpha}}\,>\,1$.
\item The position of the steepest velocity gradient is coincident with the peak in the 2D velocity dispersion map, within the uncertainties.
\end{enumerate}

Aside from the inclination-correction for $V_{rot}$, these first three criteria rely only on information from the spatially-resolved kinematic maps. \citet{2015ApJ...799..209W} report that 68\% of star-forming galaxies at $z\sim2$ satisfy these criteria. The last two criteria include:

\begin{enumerate}
\setcounter{enumi}{3}
\item The photometric and kinematic axes are aligned within 30 degrees.
\item The centroid of the continuum center is coincident, within the uncertainties, with the position of the steepest velocity gradient.
\end{enumerate}

We will also explore one additional criterion:

\begin{enumerate}
\setcounter{enumi}{5}
\item The {\emph{HST}} H-band image contains only a single nucleus.
\end{enumerate}

\begin{figure*}
\begin{centering}
\includegraphics[angle=0,scale=0.58]{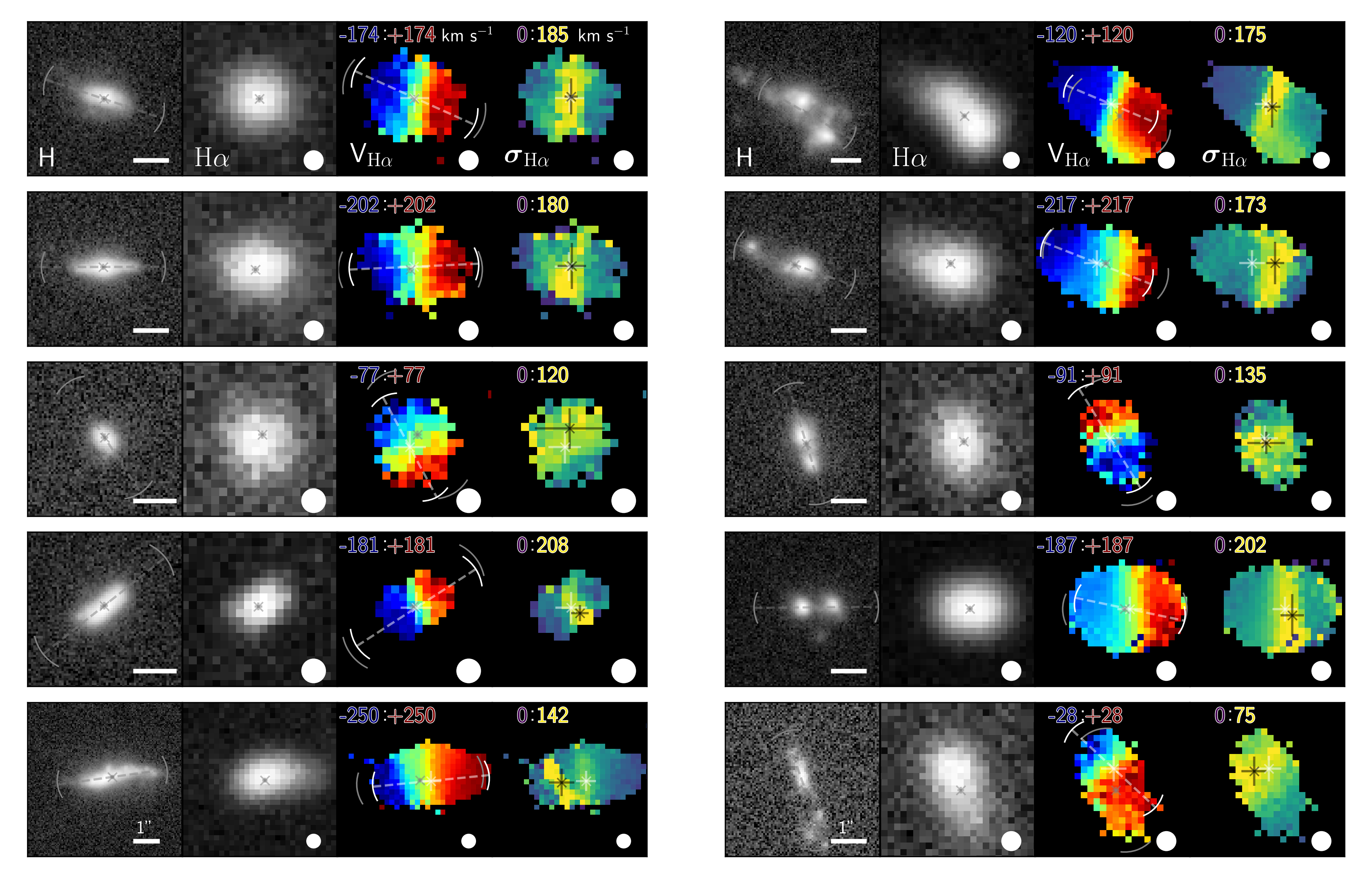}

\caption{For isolated disk galaxies (left panel) and merging systems without disks (right panel), the following are shown: synthetic {\emph{HST}}/WFC3 H-band images (first column), VLT/KMOS H$\alpha$ maps (second column), and VLT/KMOS velocity and velocity dispersion maps (third and fourth column). The intrinsic states are determined directly from the young star particles, as described in Figure {\ref{fig:images_intrinsic1}}. If not sufficiently resolved, the orbital motions of merging galaxies and the rotational signatures of disks can produce identical observational signatures: a smooth $H{\alpha}$ flux map, a smooth and monotonic velocity gradient, a velocity difference across the map that exceeds the average velocity dispersion, a dynamical center of the velocity map that is coincident within errors with a peak in the velocity dispersion map, a continuum centroid that is coincident within errors with the dynamical center of the velocity map, and an alignment between the photometric and kinematic major axes. Without considering the disturbed morphologies or double nuclei of the {\emph{HST}} images, the merging systems in the right panel would be classified as disks. The following quantities are indicated: ground-based seeing used for the synthetic KMOS spectral cube (white circle), the centroid of the continuum {\emph{HST}} image (grey x), the dynamical center of the velocity map (white x), and the peak in the velocity dispersion map (black x). The numbers listed in the velocity and velocity dispersion maps indicate the minimum and maximum of the color scale used. The photometric and kinematic major axes are indicated by grey and white lines, respectively, and 30$^{\circ}$ arcs. The arcs overlap in all snapshots, indicating that these axes are aligned. \vspace{0.1cm}}
\label{fig:many_cases}
\end{centering}
\end{figure*}

\section{Confusing Mergers for Disks in Resolved Kinematic Maps}

\subsection{Synthetic Observations of a Single Simulated Merger}
We first examine the synthetic observations of a single simulation, namely VELA 20. The central galaxy in the simulation box merges with a slightly less massive companion galaxy at z = 1.6. Prior to their encounter, the stellar mass of the central and companion are 4.1 $\times10^{10}$ M$_{\odot}$ and 1.2 $\times10^{10}$ M$_{\odot}$, respectively, qualifying this as a major merger. We follow the merger over 4 timesteps, each separated by $150$ Myr. The angular momentum profile, synthetic {{\emph{HST}}/WFC3 H-band image, synthetic VLT/KMOS H$\alpha$ flux map, and synthetic KMOS H$\alpha$ velocity and velocity dispersion maps are shown for each timestep in Figure \ref{fig:single_case}.

Over the first two snapshots, $1.6<z<1.7$, the central galaxy has a rotationally-supported star-forming disk (Figure \ref{fig:single_case}; top left) --- the young star particles are distributed near $j_z/j_{circ}=1$ and extend out to $15$ kpc from the galaxy center. As expected, the H$\alpha$ velocity and velocity dispersion maps reflect this ordered rotation (Figure \ref{fig:single_case}; bottom left) --- the primary velocity gradient is smooth and continuous, the kinematic and photometric axes are aligned, the velocity dispersion map peaks near the center of the steepest velocity gradient, and the inclination-corrected rotation velocity (345 $\pm$ 11 km s$^{-1}$) exceeds the typical velocity dispersion (61 $\pm$ 7 km s$^{-1}$). This galaxy satisfies the observational disk criteria during these time steps, illustrating their success when applied to disk galaxies with no nearby companions.

The observations are significantly more challenging to interpret during the first interaction at $z=1.56$. Before and during the merger, the central galaxy retains its rotationally-supported disk and the orbital direction of the merger is aligned with the rotational direction of the disk (top panel). At the time of collision, the maximum velocity of the velocity map, i.e., the most redshifted component, traces the rotational signature of one side of the disk, while the minimum velocity of the velocity map, i.e., the most blueshifted component, traces the orbital motion of the companion galaxy. When taken together, these two define the kinematic major axis, i.e., the direction between the maximum and minimum velocities. Instead of tracing the rotational motion of the disk --- which is still present in the map --- the kinematic major axis follows the velocity gradient defined by one edge of the disk and the companion galaxy. This measurement has significantly less physical meaning than either the rotation velocity of the disk or the relative velocity between the two galaxies.

At face-value, the velocity map is consistent with rotational motions --- the velocity gradient between the maximum and minimum velocity is smooth and continuous and velocity dispersion peaks along the steepest observed gradient.

The line-of-sight velocity difference between the edge of the disk and the merger is 546 km s$^{-1}$, while the line-of-sight velocity difference of the disk is 358 km s$^{-1}$. As such, a measurement of $V_{rot}\times \sin i$ along the kinematic major axis overestimates the true rotation velocity of the disk by a factor of 1.5 or 0.18 dex. 

Figure \ref{fig:vela20_histogram} shows the distribution of the observed inclination-corrected rotation velocities, as measured from the kinematic major axis, for 16 sightlines around VELA 20 at $z = 1.56$. For the majority of sightlines, 13/16, the derived rotation velocity along the kinematic major axis is higher than both the intrinsic rotation velocity of the central disk and the maximum circular velocity --- the latter being unphysical.

In this example, the galaxy is sufficiently large, $20$ kpc or $2.''5$, compared to the size of the ground-based seeing, $0.''6$, that the complicated kinematic signatures of the merger are apparent in the velocity maps, specifically that there are two distinct velocity gradients in the map. As follows, we examine cases where the galaxy is less resolved in the seeing-limit.

\begin{figure}
\begin{centering}
\includegraphics[angle=0,scale=0.60]{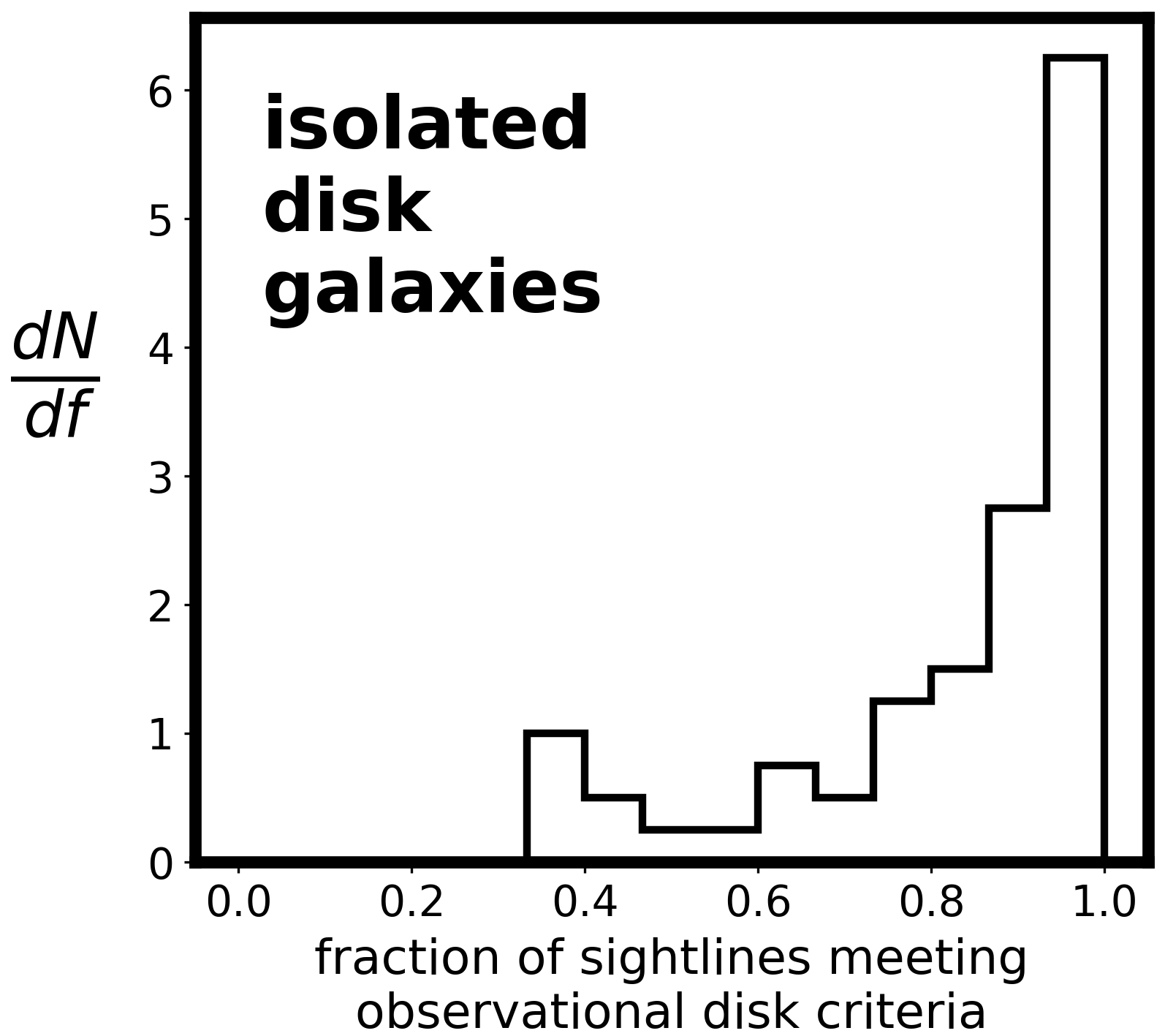}
\caption {For simulated isolated galaxies with young stellar disks, we measure the fraction of sightlines that pass the observational disk criteria. This histogram shows their distribution peaks at one, indicating that the observational disk criteria are reliable for isolated disk galaxies. \vspace{0.1cm}}
\label{fig:isolated_hist}
\end{centering}
\end{figure}

\subsection{Mergers Passing as Disks}

We now examine the full suite of simulations and consider only two cases: those snapshots in which the central galaxy has no disk and a nearby companion and those snapshots in which the central galaxy has a disk and is isolated. For the purposes of analyzing clean and distinct samples, other scenarios (i.e., those snapshots with a disk and a companion or with a non-disk that is isolated) are ignored.

We show synthetic data for example isolated disk galaxies from the VELA simulations in the left panel of Figure \ref{fig:many_cases}. The central galaxies in all of these examples exhibit regular morphological and kinematic properties. More specifically, they satisfy the observational disk criteria outlined in \S5.3 and in \citet{2015ApJ...799..209W}. 

We consider all the isolated galaxy snapshots with a rotationally-supported young stellar disk in the VELA simulations, and find that the observational disk criteria work with high confidence. In Figure \ref{fig:isolated_hist}, we show the distribution of the fraction of sightlines in which an isolated intrinsic disk is classified as a disk. This distribution is highly skewed towards a value of 1, indicating that, in a majority of sightlines, isolated disks are correctly classified as disks in the synthetic observations. 

There is, however, a shallow tail in the distribution, indicating that a non-negligible fraction of isolated intrinsic disks are misclassified as non-disks. This is most often due to the misalignment of photometric and kinematic axes and/or the offset between the continuum center and the dynamical center (i.e., failure of criteria 4 and 5). In these cases, the young star-forming regions are structurally dissimilar to the bulk of the stars --- the former determining the kinematic morphology and the latter determining the continuum morphology.

We present synthetic data for examples in which galaxies are dispersion-supported and have a nearby companion in the right panel of Figure \ref{fig:many_cases}. Although these galaxies do not intrinsically have disks, the synthetic data exhibit many of the same characteristics that we expect for disks. In all of the examples presented, two of the first three criteria outlined in \S5.3 are satisfied --- there is a smooth and continuous velocity gradient and the observed inclination-corrected ``rotation-velocity" (in actuality, the orbital motion of the merger) exceeds the velocity dispersion (as measured at the galaxy outskirts). The third criterion --- that the dynamical center is coincident with a peak in the velocity dispersion --- is satisfied in all but the second row. The fourth criterion --- that the photometric major axis is aligned with the kinematic major axis --- is satisfied in all rows. The fifth criterion --- that the center of the continuum is coincident within the uncertainties of the dynamical center --- is satisfied in all but the fifth row. 

However, it is also apparent from the high-resolution {\emph{HST}} H-band image that these galaxies are either highly disturbed or, as we discuss below, clearly have two nuclei.

These examples illustrate the challenge in distinguishing disks and mergers in seeing-limited data. Isolated disk galaxies are typically identified as disks using the observational disk criteria, but so too are mergers and sufficiently close pairs.

\subsection{Contamination as a Function of 3D Pair Separation}

In Figure \ref{fig:radius_merger}, we show the fraction of sightlines meeting different subsets of the disk criteria for each snapshot of each galaxy. We consider only those galaxies that are dispersion-supported ($\left<j_z/j_{circ}\right>_{young} < 1/3$) and with a nearby companion. The fraction is shown as a function of the 3D separation between the two galaxies.

In the top panel, we show the fraction of galaxies meeting the first 3 disk criteria outlined in \S5.3. Between $10-100\%$ of the sightlines in these galaxies --- with no intrinsic disk --- are classified as disks from these criteria. We note that there is only a weak apparent dependance on the projected separation of the central and companion galaxies. 

These first 3 criteria only use information from the kinematic maps. When a merger is misclassified as a disk, it is primarily due to the orbital motion of the merging galaxies. This orbital motion will often determine the direction of the kinematic major axis, while still satisfying the criteria $V_{rot}/\sigma_g>1$ --- where $V_{rot}$ is, in reality, measuring the velocity difference of the two galaxies. The central peak in the velocity dispersion results from the steep velocity step-function midway between the two galaxies, i.e., beam smearing of the orbital motions.

The situation improves when we consider the photometric criteria. In the middle panel, we include the 4th and 5th disk criteria --- the photometric and kinematic axes are aligned within 30 degrees and the steepest velocity gradient is coincident within the errors of the centroid of the continuum center. Adding these two criteria reduces the fractions to between 0 and 50\%. 

Finally, in the bottom panel, we explore an additional 6th criterion \S5.3 --- that the {\emph{HST}} image contains only a single nucleus. We note that when high-resolution ancillary imaging and spectra are available, several studies adopt a similar criterion to select against merging galaxies. This is typically carried out through visual inspection or through quantitative and objective measures such as projected distances, redshift separations, and mass ratios (e.g., \citealt{2017ApJ...842..121U}). This final criterion dramatically improves upon the previous 5. The majority of non-disk galaxies are correctly identified as non-disks, except in two cases where the galaxies are $<5$ kpc average projected separation.

\begin{figure}
\begin{centering}
\includegraphics[angle=0,scale=.60]{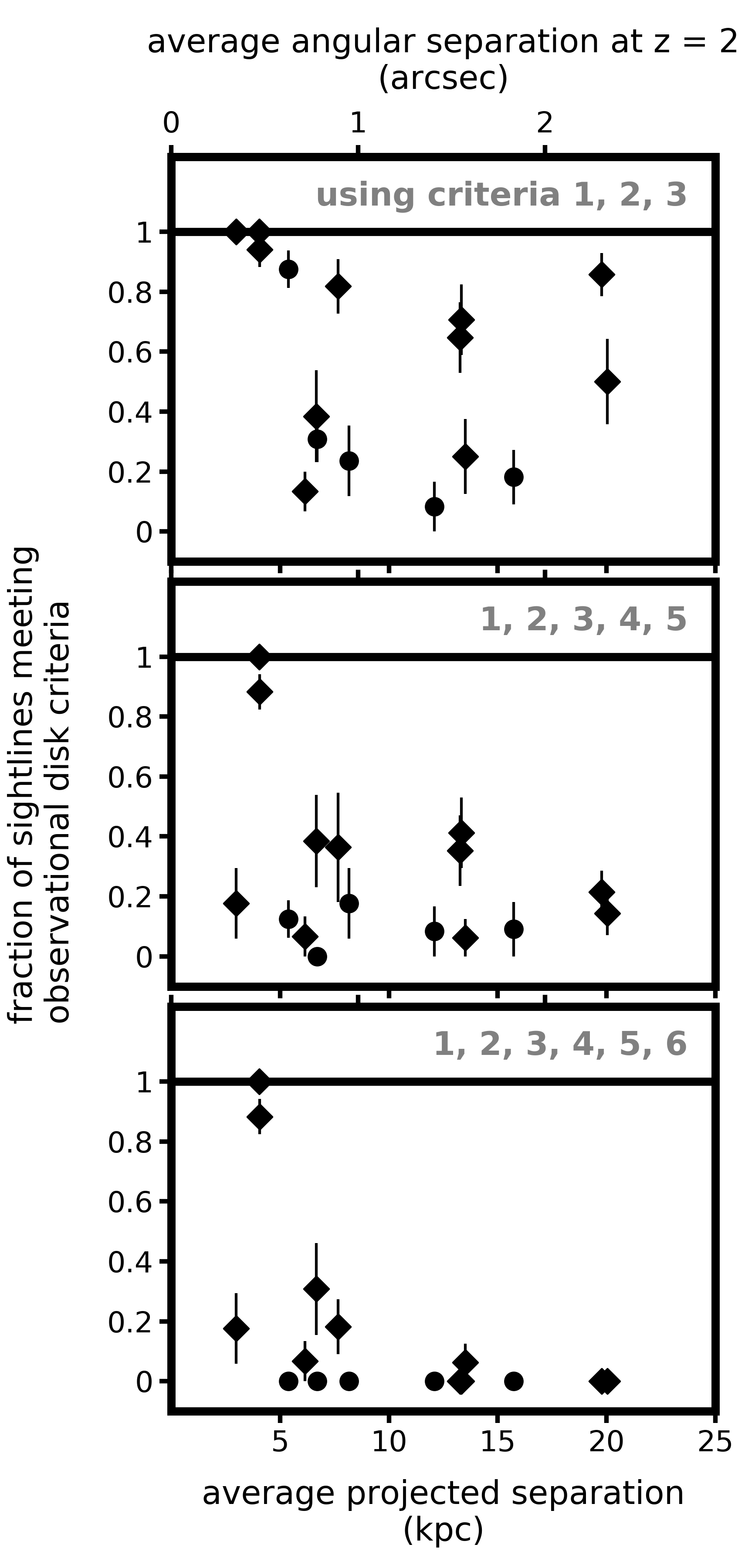}
\caption{We examine the efficacy of two sets of disk criteria commonly used by observers (top two panels), and the effect of adding an additional criterion (bottom panel). The criteria are applied to synthetic observations of merging galaxies without disks. Each panel shows the fraction of sightlines which meet the disk criteria versus the average projected separation of the merging galaxies. Each point represents a snapshot of a simulation. Ideally, the fraction of sightlines indicating a disk should be zero. However, this is not the case. The fraction of sightlines for which a major merger passes as a disk ranges from 0 to 100$\%$, depending strongly on the criteria adopted, and weakly on the separation of the merging galaxies. Uncertainties on the fractions are calculated by bootstrap resampling. Major mergers (stellar mass ratios 1:1 -- 1:4) and minor mergers (stellar mass ratios 1:4 -- 1:30) are shown as circles and diamonds, respectively. These fractions are used later to estimate the effect of mergers on observed disk fractions.}
\label{fig:radius_merger}
\end{centering}
\end{figure}

\begin{figure*}
\begin{centering}
\includegraphics[angle=0,scale=.75]{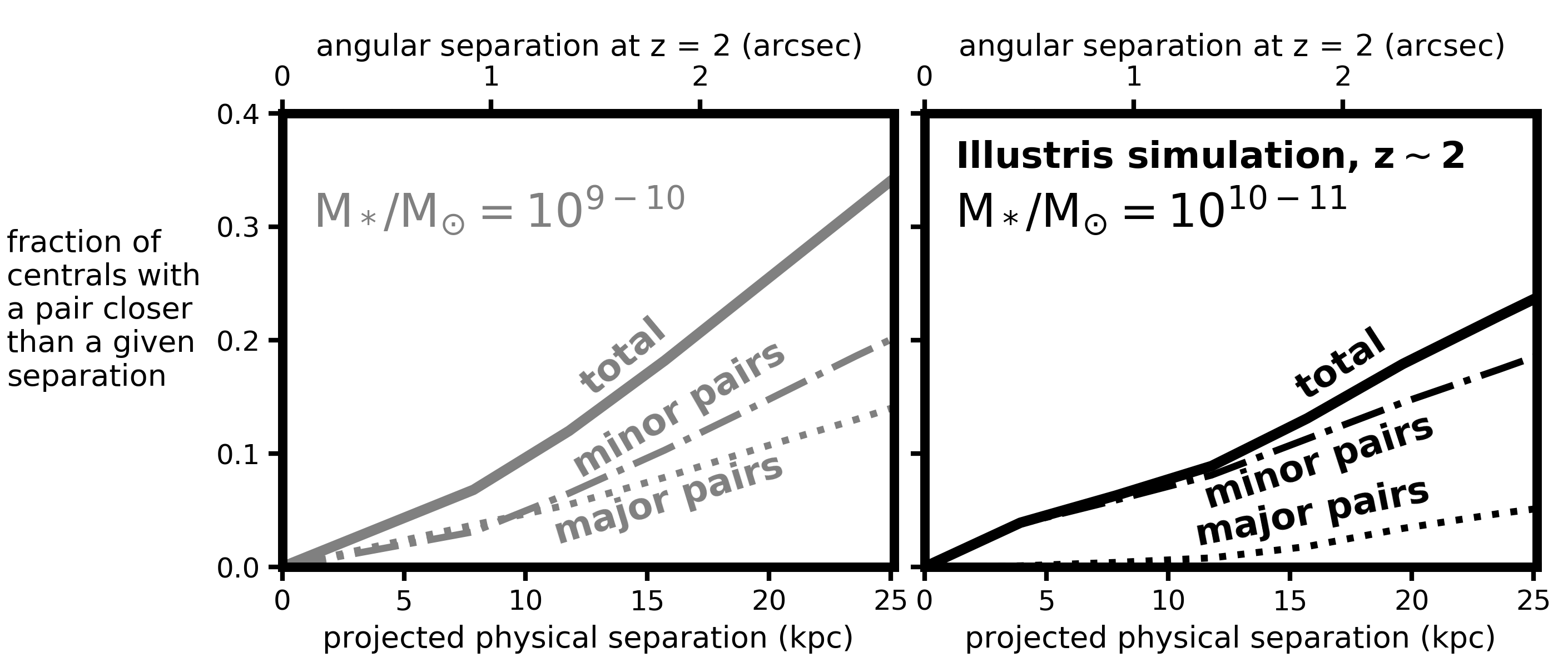}
\caption {To estimate the fraction of galaxies undergoing a merger at $z\sim2$, we use close pair fractions from the Illustris simulation. The fraction of Illustris galaxies at $1.5 < z < 3.0$ with at least one companion closer than a given projected separation is shown for two bins in stellar mass: $10^{9-10}$M$_{\odot}$ (left) and $10^{10-11}$M$_{\odot}$ (right). These fractions will be combined with the results of  Figure {\ref{fig:radius_merger}} to calculate corrections to the observed disk fractions for mergers passing as disks. The stellar mass bins shown the typical stellar mass ranges of high redshift kinematic samples in the literature. Two bins in close pair mass ratio are shown -- 1:1 - 1:4 (major pairs; dotted line) and 1:4 -- 1:30 (minor pairs; dashed-dotted line).}
\label{fig:illustris_pairfractions}
\end{centering}
\end{figure*}

\section{Quantifying the Effect of Mergers on Measurements of the Global Disk Fraction at High Redshift}

In this next section, we use theoretically-derived pair fractions to estimate the contamination of mergers to the observed fraction of disks at high redshift ($1\lesssim z \lesssim 3$).

We have demonstrated that, due to seeing and small galaxy sizes, galaxies at $z\sim2$ with a nearby companion and no disk can display the same kinematic signatures as those expected from disks (i.e., those outlined in \S5.3).  The fraction of sightlines that an observer will incorrectly identify a merger as a disk depends on both the observables available and the physical separation of the merging galaxies --- ranging between 10 -- 100\% if only kinematic information is available and 0 -- 50\% if both kinematic and photometric information is available (Figure \ref{fig:radius_merger}).

To estimate the degree of this effect on real observations, one must know the fraction of galaxies with companions or late-stage mergers --- i.e., those at a projected separation $<\,$25 kpc --- as a function of mass and redshift. 

Through real observations of high redshift galaxy pairs, \citet{2018MNRAS.475.1549M} find that the fraction of massive galaxies (M$_*$/M$_{\odot}\,>2\times10^{10}$) with major nearby companions ($4:1$ mass ratio; $5-50$ kpc projected separation) increases with redshift, to $\sim$$15\%$ at $z\sim1$, before beginning to decline. This is consistent with the close pair fractions computed by \citet{2017MNRAS.465.1157R} ($\sim 15-17\%\,\pm\,6\%$, depending on the stringency of selection choice) at $z\sim 1$, once the close-pair selection cuts employed by both studies are made consistent.

As an important aside --- given that the kinematic maps are luminosity-weighted, the appropriate statistic is the {H$\alpha$}-flux pair fraction--- i.e., companion galaxies that have sufficiently high H$\alpha$ flux to register velocity values in otherwise unregistered spaxels. It is not straightforward to convert between a mass ratio-selected pair fraction and an H$\alpha$ ratio-selected pair fraction. For instance, \citet{2018MNRAS.475.1549M} find that $\sim$$20\%$ of galaxies qualify as a major pair at $z\sim2.5$ when selecting by H-band flux ratio (4:1 flux ratio), while only $5\%$ qualify when selecting by stellar mass ratio (4:1 mass ratio).

\subsection{The Fraction of Central Galaxies with a Companion in the Illustris Simulation}
In lieu of using observed pair fractions, which are subject to incompleteness at low stellar masses and/or high mass ratios, we use results from the (106.5 Mpc)$^{3}$ Illustris cosmological galaxy formation simulation box \citep{2014MNRAS.445..175G, 2014MNRAS.444.1518V}. Illustris mergers have been studied in detail by \citet{2015MNRAS.449...49R}. Pair fractions are calculated by \citet{2017MNRAS.468..207S} using lightcones, a technique for generating realistic synthetic surveys from cosmological simulations. Specifically, they followed \citet{2007MNRAS.376....2K} to generate three 140 square arcmin lightcone catalogs from which to select pairs using common observational criteria. Using simulation results here has its own limitations owing to their uncertain galaxy formation physics, and the challenge of properly assigning mass at small galaxy-galaxy separation (see extensive discussion in \citealt{2015MNRAS.449...49R}), but the simulated pair statistics results agree reasonably well with data in the regions of parameter space where they overlap \citep{2017MNRAS.468..207S}.

For consistency with the rest of the paper, we first select pairs using the 3D separation of galaxies in the synthetic Illustris lightcone catalog. Then, using the results of the Appendix, we convert the 3D pair fraction statistics to the more observationally-accessible 2D statistics --- pair fraction as a function of average projected 2D separation. 

Figure \ref{fig:illustris_pairfractions} shows cumulative pair fractions as a function of projected separation on the sky at $1.5\,<\,z\,<\,3$ from the synthetic Illustris lightcone catalog. These are broken into two bins of stellar mass --- $10^{9}\,<\,$M$_{*}$/M$_{\odot}<10^{10}$ (left panel) and $10^{10}<\,$M$_{*}$/M$_{\odot}<10^{11}$ (right panel). We further split these bins into two bins in stellar mass ratio (i.e., stellar mass of the companion: stellar mass of the central) --- 1:1 -- 1:4 (i.e., major pairs), 1:4 -- 1:30 (i.e., minor and very minor pairs).

\subsection{Estimating the Contamination to Observed Disk Fractions from Mergers in the Pair Stage}

The observed fraction of galaxies with disks ($f_{obs}$) is related to the true fraction of galaxies with disks ($f_{int}$), the fraction of central galaxies with a sufficiently massive companion ($f_{pair}$), and the probability that, given a random sightline, a merger will pass as a disk ($X_{contamination}$):

\begin{equation}\label{eq:2}
f_{obs} = (1 - f_{int}) \times f_{pair} \times X_{contamination} +  f_{int}
\end{equation}

We determined that $X_{contamination}$ spans the full range between 0.0 and 1.0 for projected separations less than 25 kpc, depending strongly on the specific criteria used and weakly on the separation of the merging galaxies (Figure \ref{fig:radius_merger}). For simplicity, we ignore that an isolated disk can be observed as a non-disk in a small fraction of sightlines (as found in \ref{fig:isolated_hist}). We use the Illustris lightcone pair fractions to determine $f_{pair}$ (Figure \ref{fig:illustris_pairfractions}).

In Figure \ref{fig:corrections}, we use the analytic model of Eq. \ref{eq:2} to translate from observed disk fractions to true disk fraction at $z\sim2$. We adopt three values for $X_{contamination}$ that span the scatter in Figure \ref{fig:radius_merger}: 10\% (dashed-dotted line), 40\% (solid line), and 70\% (dashed line). The correction is smaller for high mass galaxies (M$_*$ = $10^{10-11}$ M$_{\odot}$) than it is for low mass galaxies (M$_*$ = $10^{9-10}$ M$_{\odot}$). 

To estimate how large the correction is for real observations, we consider observed disk fractions from both slit-based (\citealt{2017ApJ...843...46S}) and IFS-based surveys \citep{2015ApJ...799..209W}. We note that the slit-based measurement identify disks using only criteria 2 and 3 (and need to assume 1 and 4), but are in general agreement with the IFS-derived fractions \citep{2017ApJ...843...46S}. The observed value of $f_{obs}$ at $z\sim2$ is $\,\simeq\,70\%$ \citep{2015ApJ...799..209W, 2017ApJ...843...46S} for high mass galaxies and the difference between $f_{obs}$ and $f_{int}$ is $<\,5\%$. For low mass galaxies, the observed disk fraction at $z\sim2$ is $f_{obs}\,\simeq\,50\%$ \citep{2017ApJ...843...46S}, and the difference is $\,<15\%$.

In Figure \ref{fig:corrections_extreme}, we estimate the true disk fraction given observed disk fractions from \citet{2017ApJ...843...46S} with the {\emph{most extreme assumption one can make about merger contamination}} ---  that 100\% of pairs inside some projected separation ($D_{contamination}$) masquerade as disks. We explicitly show the observed value of the disk fraction, with their uncertainties (dashed line and shaded region). The true disk fraction derived from Eq. \ref{eq:2}, i.e., $f_{int}$, is shown as a solid line.

Even with this extreme assumption, the low mass galaxies only need a significant correction factor ($25-50\%$) to $f_{obs}$ for $D_{contamination}$ = 15 - 25 kpc. However, these cases almost always show two nuclei in the {\emph{HST}} imaging and should identified in samples if imaging is available (Figure \ref{fig:radius_merger}). At high mass, the correction factor is still minimal ($<5\%$), even for $D_{contamination} = 25$ kpc.

The discrepancy between the true disk fraction and the observed disk fraction due to merger contamination is minimal for both low-mass galaxies ($10^{9}\,<\,\,$M$_{*}$/M$_{\odot}\,<\,10^{10}$) and  high-mass ($10^{10}\,<\,\,$M$_{*}$/M$_{\odot}\,<\,10^{11}$) galaxies. Taking the Illustris pair fractions at face-value, the difference between the observed disk fraction and the true disk fraction is estimated to be $\lesssim5\%$ and  $\lesssim15\%$ at high and low mass, respectively.

\begin{figure}
\begin{centering}
\includegraphics[angle=0,scale=0.61]{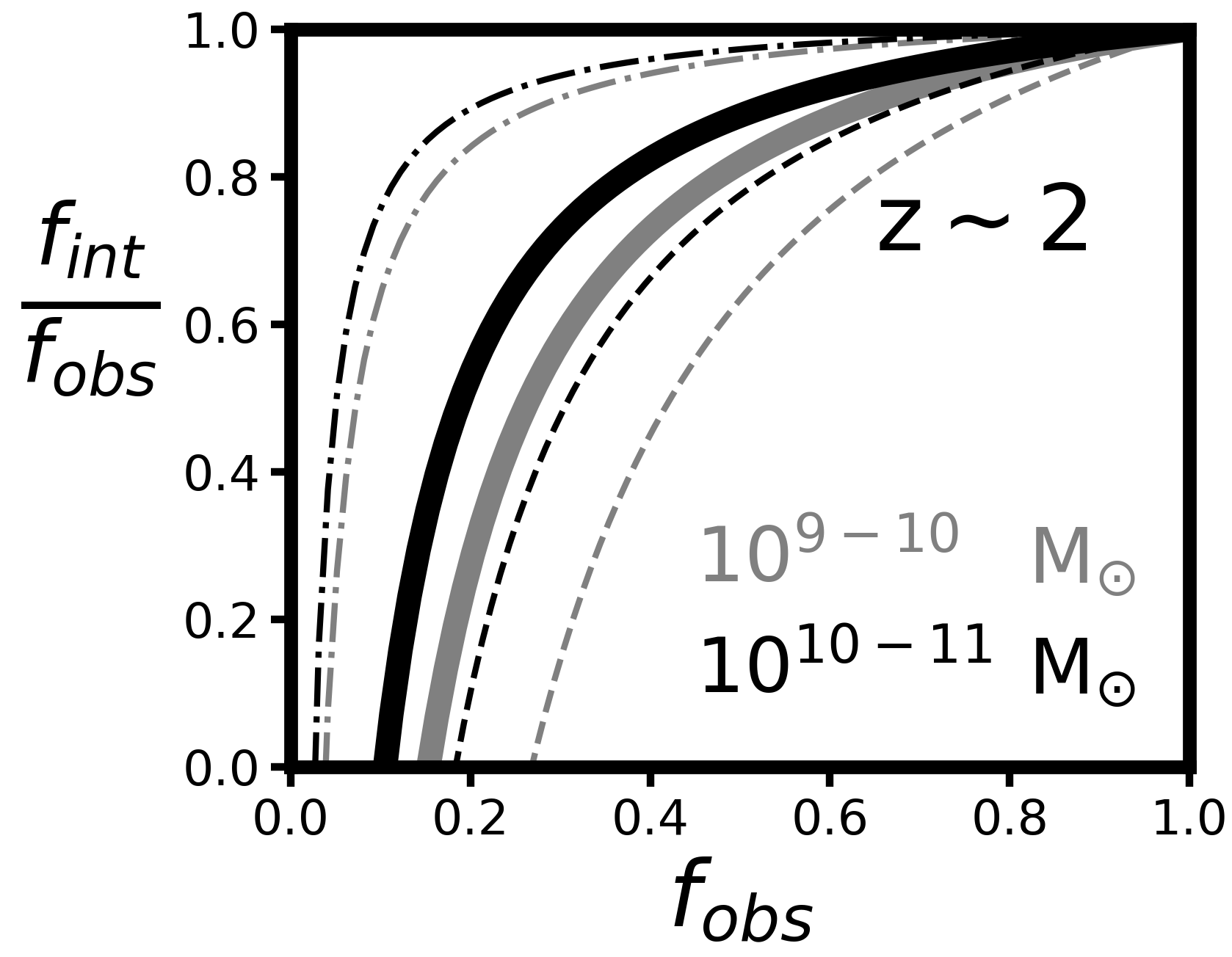}
\caption{This figure shows the implications of this paper for observations of disk fractions. Specifically, it shows how to translate observed disk fractions $f_{obs}$ to intrinsic disk fractions $f_{int}$. It is the result of a simple model which combines the fraction of galaxies in a merger  at $z\sim2$ from the Illustris simulation (Figure \ref{fig:illustris_pairfractions}), with the fraction of sightlines in which a merger will pass for a disk (Figure \ref{fig:radius_merger}). For the latter, we adopt three values that span the scatter in Figure \ref{fig:radius_merger}: 10\% (dashed-dotted line), 40\% (solid line), and 70\% (dashed line). The difference between f$_{int}$ and f$_{obs}$ is smaller for high mass galaxies (black; $10\,<\,\log\,$M$_{*}/$M$_{\odot}\,<\,11$) than it is for low mass galaxies (grey; $9\,<\,\log\,$M$_{*}/$M$_{\odot}\,<\,10$). Considering the values of $f_{obs}$ in the literature, $\sim70\%$ for high mass galaxies and $\sim50\%$ for low mass galaxies, the difference between $f_{obs}$ and $f_{int}$ is minimal ($<5\%$ and $<15\%$, for the respective mass bins).}
\label{fig:corrections}
\end{centering}
\end{figure}

\begin{figure}
\begin{centering}
\includegraphics[angle=0,scale=0.61]{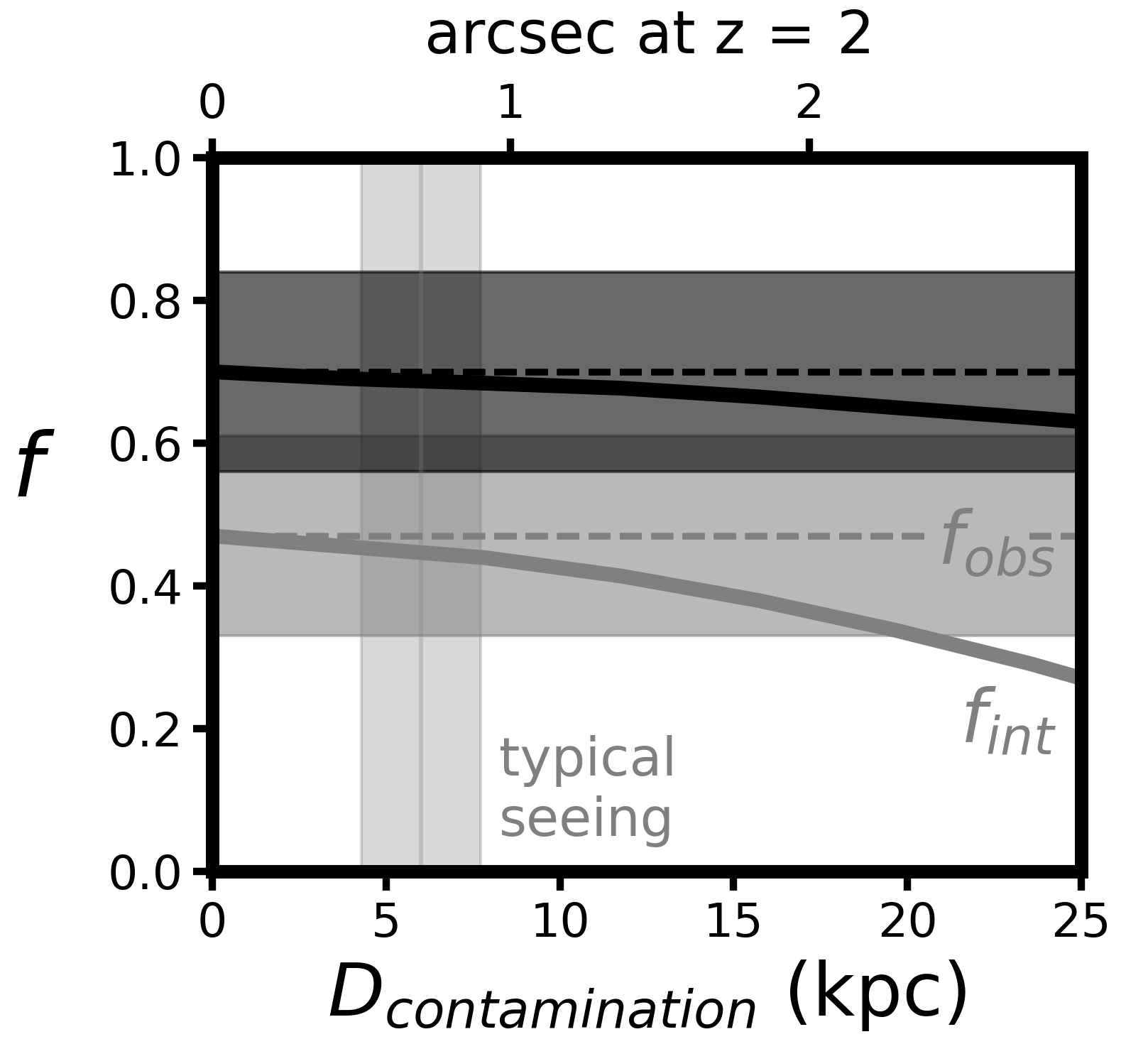}
\caption{This figure shows the difference between the observed disk fraction $f_{obs}$ (dashed line) and the true disk fraction $f_{int}$ (solid line) if we consider the worst-case scenario in our analytic model --- that 100$\%$ of mergers separated by a projected distance less than $D_{contamination}$ will pass as a disk. The difference is shown as a function of $D_{contamination}$. The values of $f_{obs}$ at $z\sim2$ are taken from \citet{2017ApJ...843...46S} and are indicated with dashed lines. Their uncertainties are indicated by shaded regions. The difference between the observed disk fraction and true disk fraction is $\lesssim15\%$ for both mass bins to $D_{contamination} = 15$ kpc or, equivalently, $\sim2\arcsec$ at $z=2$. This indicates that, even with this extreme assumption, there are not enough galaxies in mergers at $z\sim2$ to significantly impact observed disk fractions. High mass galaxies ($10\,<\,\log\,$M$_{*}/$M$_{\odot}\,<\,11$) and low mass galaxies ($9\,<\,\log\,$M$_{*}/$M$_{\odot}\,<\,10$) are shown in black and grey, respectively. The typical ground-based seeing for kinematics observations is shown with a grey swath. \vspace{0.3cm}}
\label{fig:corrections_extreme}
\end{centering}
\end{figure}

\subsection{Estimating the Contamination to Observed Disk Fractions from Mergers in the Late-Stage}

In the two subsections above, we focus on the pair and close pair stages --- where the two galaxies have a projected separation on the sky of at least 5 kpc. This owes to the uncertainty of the Illustris-derived pair fractions when the simulated galaxies are near-coalescence and overlapping \citep{2015MNRAS.449...49R}.

We now estimate the fraction of central galaxies in the late-stage/pre-coalescence stage, i.e., the mergers that are too close to be included in the pair fractions but are not yet relaxed. The fraction of galaxies in the pair stage $F_{pair}$ ($5 < d < 50$ kpc) and the fraction of galaxies in the late stage $F_{late}$ ($d < 5$ kpc) are related by their relative timescales:
\begin{equation}\label{eq:3}
F_{late} = F_{pair} \times \left(\frac{\tau_{late}}{\tau_{late} + \tau_{pair}}\right)
\end{equation}
where $\tau_{pair}$ and $\tau_{late}$ are the time a single merger will spend in the pair stage and the late stage, respectively.

The time for a merger to pass from the pair stage to the late-stage/coalescence stage, i.e., $\tau_{pair}$, was measured in Illustris as $\sim700$ Myr at $z\sim2$ \citep{2017MNRAS.468..207S}. 

The time for a merger to dynamically relax after entering the late-stage is an unknown, but we use the following simple argument to estimate its value. The kinematics observations trace the ionized gas in galaxies. As gas is collisional, it should dynamically relax within a $\sim$ few crossing times ($\sim$100 Myr) after the mergers enter the late-stage.

Let us consider $\tau_{late} = 350$ Myr and $\tau_{pair} = 700$ Myr. To associate the timescale $\tau_{pair}$ with the correct pair fraction in Eq. \ref{eq:2}, we must use the pair fraction as measured at the average separation of all pairs in the pair stage. The Illustris-derived pair fractions at $z\sim2$ ($F_{pair}$) --- the fraction of galaxies with a pair within 25 kpc projected separation --- are 20\% at high mass and 35\% at low mass (Fig. \ref{fig:illustris_pairfractions}). Given these values, $F_{late}$ as calculated from Eq. \ref{eq:3} is only 7\% and 12\%, respectively. We then use Eq. \ref{eq:2} to estimate the correction to the observed disk fraction. Even if we assume that 100\% of these late-stage mergers masquerade as disks (i.e., the most extreme assumption), the correction to the observed disk fractions is only $\sim1.8\%$ and 5.5$\%$ for high and low mass galaxies, respectively (i.e., negligible).

\section{Conclusions}

A majority ($\sim70\%$) of massive star-forming galaxies at $z\sim2$ are observed to have the following disk-like qualities: continous velocity gradients, rotation velocities that exceed their average local ionized gas velocity dispersions, central peaks in their velocity dispersion maps, and aligned kinematic and photometric major axes \citep{2015ApJ...799..209W}. It is generally inferred that these galaxies are disk-like, albeit with large amounts of disordered motions. However, with typical ground-based seeing, the orbital motions of merging galaxies can appear regular and disk-like in resolved kinematic maps \citep{2015ApJ...803...62H, 2016ApJ...816...99H}. Given the increased merger activity at $z\sim2$ compared to the local Universe, it is important to quantify the degree to which mergers are misclassified as disks in current high redshift samples.

This paper has two main conclusions. First, from seeing-limited kinematic data and Hubble images, one cannot tell with certainty whether any individual galaxy at $z\sim2$ is a merger or a disk. Second, the estimated merger rates at $z\sim2$ are not high enough for mergers to have a significant effect on measured disk fractions.

To come to these conclusions, we created synthetic {\emph{Hubble}} and KMOS IFU observations of cosmological galaxy formation simulations  at $z\sim2$, and measured the probability that a merger will pass as a disk as a function of merger stage. We found that merging galaxies pass the disk criteria in a non-negligible fraction of sightlines. This fraction ranges from 0 to 100$\%$ and depends strongly on the specific disk criteria adopted, and weakly on the separation of the merging galaxies (Fig. \ref{fig:radius_merger}).

To estimate how often mergers pass as disks in current observational samples, we combined these fractions with estimates for the number of galaxies in a merger at $z=2$ from the Illustris simulation. For galaxies with stellar masses above $10^{9}$ M$_{\odot}$, the merger rates are not high enough for this effect to have a significant impact on measurements of disk fractions (Fig. \ref{fig:corrections}). Taking the simulations at face-value, the observed disk fractions are overestimated by at most $15\%$ and $5\%$ at low  ($10^{9-10}$ M$_{\odot}$)  and high ($10^{10-11}$ M$_{\odot}$) stellar mass, respectively.

 \section*{Acknowledgements}
 
RCS would like to acknowledge support for this project through a STScI/DDRF grant to SAK. GFS appreciates support from a Giacconi Fellowship at the Space Telescope Science Institute, which is operated by the Association of Universities for Research in Astronomy, Inc., under NASA contract NAS 5-26555. KBM acknowledges support from the HST archival research grant HST-AR-15040. This work was partly supported by the US-Israel BSF grant 2014-273 (AD, JRP) and by the NSF AST-1405962 (AD). The Flatiron Institute is supported by the Simons Foundation. DC has been partly funded by the ERC Advanced Grant, STARLIGHT: Formation of the First Stars (project number 339177). NM acknowledges support from the Klauss Tschira Foundation through the HITS-Yale Program in Astrophysics (HYPA). We thank the referee for a useful report. The simulations were performed at NASA advanced Supercomputing (NAS) at NASA Ames Research Center and at the National Energy Research Scientific Computing Center (NERSC), Lawrence Berkeley National Laboratory. This work has made us of the publicly available software {\tt{YT}} \citep{2011ApJS..192....9T}. This research made use of Astropy, a community-developed core Python package for Astronomy \citep{2013A&A...558A..33A}. 

\appendix
\section{From 3D separation to average 2D projected separation}

Throughout this paper, we substitute a physical quantity measured from the simulations --- the 3D separation between a central galaxy and a companion galaxy --- with a quantity more relevant for comparing with observations --- their average projected separation on the sky. As shown below, these quantities are related by a factor $\pi/4$. Let us define an arbitrary 3D point ${\vec{r}} = (x, y, z)$ with a 3D distance from the origin $\rho$:

\begin{align}
{\vec{r}} &= x\hat{x}+y\hat{y}+z\hat{z}\\
&=\rho\sin\phi\cos\theta\hat{x}+\rho\sin\phi\sin\theta\hat{y}+\rho\cos\phi\hat{z} \nonumber
\end{align}

The projected separation $R$ of this point and the origin for an observer normal to $\hat{z}$ is:

\begin{align}
R &= \sqrt{x^2+y^2}\\
& = \sqrt{(\rho\sin\phi\cos\theta)^2+(\rho\sin\phi\sin\theta)^2} \nonumber\\
& = \rho\,\sin\phi \nonumber
\end{align}

The average of $R$ over a sphere of fixed radius $\rho$ is:

\begin{align}
<R> &= \frac{\int^{2\pi}_0\int^\pi_0R\,d\Omega}{\int^{2\pi}_0\int^\pi_0\,d\Omega}\\
& = \frac{\int^{2\pi}_0\int^\pi_0(\rho\sin\phi)(\sin\phi\,d\phi\,d\theta)}{\int^{2\pi}_0\int^\pi_0\sin\phi\,d\phi\,d\theta}\nonumber\\
& = \rho\frac{\int^\pi_0\sin^2\phi\,d\phi}{\int^\pi_0\sin\phi\,d\phi}\nonumber\\
& = \rho\left(\frac{\pi}{4}\right)\nonumber
\end{align}

That is, the average 2D projected separation of two points is $\pi/4$ times their 3D separation.

\bibliography{RCS_thesis}{}
\bibliographystyle{apj}

\end{document}